\documentclass{camera}
\usepackage{graphicx}  

\def\c2r{\chi^2_\nu}
\def\c2{\chi^2}

\def\epo{$E_{\rm p}$ }
\def\epi{\ensuremath{E_{\rm p,i}}}
\def\eiso{\ensuremath{E_{\rm iso}}}

\def\eiso{\ensuremath{E_{\rm iso}}}

\def\epeiso{$E_{\rm p,i}$ -- $E_{\rm iso}$}
\def\nufnu{$\nu F_{\nu}$ }

\def\swift{{\it Swift}}
\def\fermi{{\it Fermi}}

\def\h0{H$_{\rm 0}$~}

\begin{document}

%
\title{The \epeiso{} Correlation and \fermi{} Gamma--Ray Bursts}

%
\author{L. Amati}

%
\organization{Italian National Institute for Astrophysics (INAF) -- IASF Bologna, 
via P. Gobetti 101,
40129 Bologna, Italy}

\maketitle

\begin{abstract}

The \epeiso{} correlation is one of the most intriguing properties of GRBs, with 
significant implications for the understanding of the physics and geometry of the 
prompt emission, the identification and investigation of different classes of GRBs, 
the use of GRBs as cosmological probes. The \fermi{} satellite, by exploiting the 
high accuracy of the GBM instrument in the measurement of \epi{}, the simultaneous 
detection of GRBs with \swift{}, and the detection and localization of GRBs in the 
GeV energy range by the LAT instrument, is allowing us to enrich the sample of of 
GRBs with known redshift and reliable estimate of \epi{} and, thus, to further test the 
robustness, reliability and extension of this correlation. Based on published 
results and preliminary spectral data available as of the end of 2009, it is found 
that the locations in the \epeiso{} plane of \fermi{} long and short GRBs with 
measured redshift, including extremely energetic events, are consistent with the 
results provided by previous / other experiments.

\end{abstract}

%
\section{Introduction}

Since 1997, the measurements of the redshift of a significant fraction of events 
(more than 200 nowadays) is giving us the possibility of performing systematic 
investigations of the intrinsic properties of Gamma--Ray Bursts (GRBs). Among 
these, the correlation between the photon energy, \epi{}, at which the \nufnu{} 
spectrum (in the cosmological rest--frame of the source) of the prompt emission 
peaks, and the isotropic--equivalent radiated energy, \eiso{}, is one of the more 
intriguing and debated \cite{Amati02,Amati06}. Indeed, several observational and 
thoretical studies have shown the relevance of the existence and properties 
(namely, the slope and the dispersion) of the \epeiso{} correlation for the models of 
the GRB prompt emission, whose physics and geometry are still to be settled 
\cite{Zhang02,Amati06,Amati08a}. Furthermore, it was found that short GRBs do not 
follow the correlation, as is true for the peculiarly sub--energetic and close 
GRB\,980425, the proto--type of the GRB/SN connection (see Figure~1). These 
evidences suggest that the \epeiso{} plane can be used to distinguish between 
different classes of GRBs and to understand the differences in the physics / 
geometry of their emission \cite{Amati06b,Amati07,Piranomonte08}. The \epeiso{} 
correlation, together with other "spectral--energy" correlation derived from it by 
adding more observables or substituting \eiso{} with the average peak luminosity, 
were also investigated as a promising tool for the estimate of cosmological 
parameters \cite{Ghirlanda06,Schaefer07,Amati08b}.

Thus, the enrichment of the sample of events with known redshift and \epi, together 
with the
investigation of the impact of selection and instrumental effects on the 
correlation, is of key importance for this field of research. Under this respect, the 
\fermi{} satellite, thanks to the unprecedently broad energy band ($\sim$8 keV -- 
$\sim$50 MeV) of its GRB Monitor (GBM) is expected to provide a significant 
contribution. This is particularly true in the present "golden era", in which the 
\swift{} satellite, thanks to its fast and accurate location of the early afterglow 
emission, is allowing prompt follow--up of GRBs with optical telescopes and, 
consequently, a significant increase of the numbe of redshift estimates. 
In addition, the 
\fermi/LAT can detect and localize those bright GRBs with emission extending up to 
the GeV energy range, thus allowing the investigation of the spectral--energetic  
properties of peculiarly energetic events and to further test the robustness and 
extension to the \epeiso{} correlation.

\begin{figure}
\centerline{\includegraphics[width=11cm]{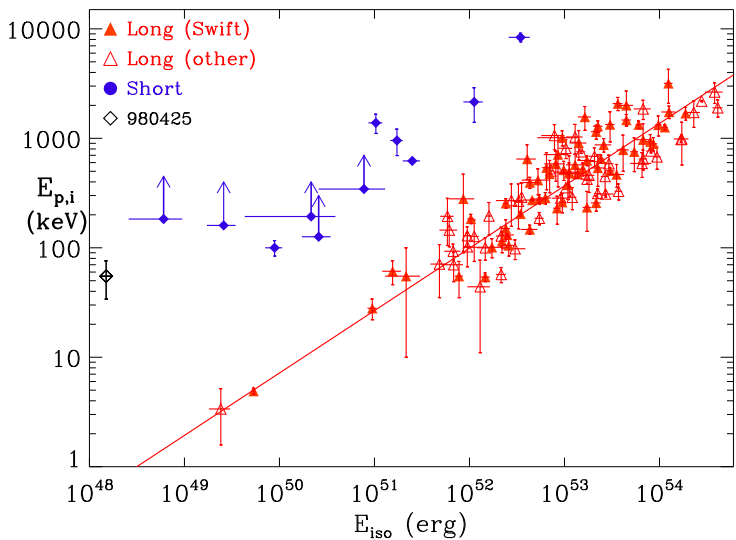}}
\caption{GRBs in the \epeiso{} plane as of end of 2009. The continuous line is the 
best fit power--law of the 108 long GRBs.}
\label{fig01} 
\end{figure}

\section{The \epeiso{} correlation: observational status}

Figure~1 shows the location in the \epeiso{} plane of those GRBs with measured 
redshift {\and} spectral peak energy, as of the end of 2009. The sample of long 
GRBs includes 108 events. For the 95 events up to April 2009, the data are taken 
from \cite{Amati08b} and 
\cite{Amati09}. For the remaining 13 
GRBs (090516, 090618, 090715B, 090812, 090902B, 090926, 090926B, 091003, 091018, 
091020, 091029, 091127, 091208B) the values of \epi{} and \eiso{} were calulated based 
on the redshift, fluences and spectral parameters reported in the GCNs 
({\it gcn.gsfc.nasa.gov}) and following the method 
described, e.g., in \cite{Amati06}. The values for short GRBs are from 
\cite{Amati06b,Amati09,Piranomonte08}. 

As can be seen, the \epi{} and \eiso{} values of all long GRBs, with the already 
mentioned exception of the peculiar GRB\,980425, are strongly correlated (Spearman's 
$\rho$ = 0.86 for 108 events). The \epeiso{} correlation, as reported also in 
previous works \cite{Amati08b, Amati09}, can be modeled with a power--law 
with index $\sim$0.5 and an extrinsic dispersion (i.e., the scatter of the data in 
excess to that expected based on Poissonian fluctuations only) 
$\sigma$({\rm log}\epi)$\sim$0.2. 
As discussed above, even if the sample is still small, there is clear 
evidence that short GRBs do not follow the correlation holding for long ones. This 
fact gives further clues on the different emission mechanisms at work in the two 
classes of events and makes the \epeiso{} plane a useful tool to distinguish between 
them \cite{Antonelli09}. A natural explanation for the short/long 
dicotomy and the different locations of these classes of events in the \epeiso{} 
plane is provided by the "fireshell" model of GRBs \cite{Ruffini09}.

The impact of selection and instrumental effects on the \epeiso{} correlation of long 
GRBs was investigated since 2005, mainly based on the large sample of BATSE GRBs 
without known redshift. Different authors came to different conclusions 
\cite{Nakar05,Band05,Ghirlanda05}. In particular, \cite{Ghirlanda05} showed 
that BATSE events potentially follow an \epeiso{} correlation and that {\it the 
proper question is not if the correlation is real but if, and how much, its 
measured dispersion is biased}. There were also claims that a significant fraction 
of \swift{} GRBs is inconsistent with the correlation \cite{Butler07}. 
However, as can be seen in 
Figure~1, when considering those \swift{} events with peak energy measured by 
broad--band instruments like, e.g., Konus--WIND or the \fermi/GBM (see next 
Section) or reported by the BAT team in their catalog \cite{Sakamoto08} or GCNs, 
it is found that they are all consistent with the \epeiso{} correlation as determined 
with previous/other instruments \cite{Amati09,Krimm09}. In addition, 
\cite{Amati09} also found that the slope and normalization of the 
correlation based on the single data sets provided by GRB detectors with different 
sensitivities and energy bands are very similar. These evidences further support 
the reliability of the correlation.

\section{\fermi{} GRBs in the \epeiso{} plane}

As discussed above, in the last five years, thanks to the unprecedented capabilities 
of the the \swift{} satellite, the fraction of GRBs with redshift estimate increased 
significatly. However, due to the limited energy band (15--150 keV) of the BAT
GRB monitor, \swift{} can estimate \epo{} only for 15--20\% of the events. Indeed, for most 
of the \swift{} GRBs that can be placed in the \epeiso{} plane (Figure 1), the peak energy
could be measured thanks to the simulatenous detection by other detectors with better 
spectral capabilities (e.g., Konus--WIND). In ths context, 
with its wide field of view and unprecedently broad energy band, 
the \fermi/GBM is expected to provide an important contribution by 
significantly increasing the number and accuracy of the estimates of \epi{} for
GRBs with measured redshift. 
 
In Figure~2, I show the location in the \epeiso{} plane of those GRBs for which the 
spectral parameters were provided by the GBM, with the only exception of GRB\, 
090323 (also detected by the LAT), for which the spectrum of the whole event was 
reported only by the Konus--WIND team \cite{Golenetskii09}.
The red dashed line is the best fit power--law of this sample, whereas the 
continuous line show the best--fit power--law and $\pm$2$\sigma$ dispersion region 
of the \epeiso{} correlation as determined by \cite{Amati08b} based on a sample 
of 75 GRBs not including \fermi{} events. As can be seen, \fermi{} GRBs follow the 
\epeiso{} correlation, with a slope and dispersion well consistent with those 
determined based on measurements by other satellites. 

Figure~2 also shows that, as already pointed out and discussed by 
\cite{Amati09}, 
the extremely energetic events with GeV emission detected and localized 
by the LAT, GRB\,080916C and GRB\,090323, are consistent with the correlation and 
extend it to higher energies. Pushed by the extension of the spectrum of GRB\,080916C
up to several GeVs without any significant deviation from
the Band function describing the soft gamma--rays emission, these authors also 
investigated the impact of the extension 
from 10 MeV up to 10 GeV of the energy band on which \eiso{} is computed, finding no
significant changes in the slope and dispersion of the \epeiso{} correlation.

Finally, \fermi{} provided the most accurate estimate of \epi{} for a short burst with
measured redshift, GRB\,090510 . The apparent deviation of this event from
the \epeiso{} correlation is a further confirmation of results obtained by
previous satellites.

It has to be cautioned that  
the spectral parameters and fluences
published in the GCNs by the GBM team are still preliminary. However, the results 
coming from a more refined analysis are not expected to be so different to 
drastically change the above conclusions.

\begin{figure}
\centerline{\includegraphics[width=11cm]{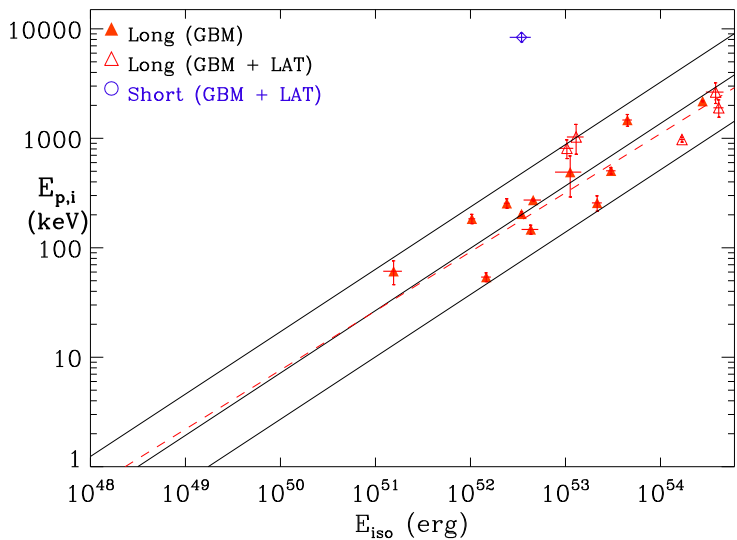}}
\caption{\fermi{} GRBs in the \epeiso{} plane as of end of 2009. The black 
continuous lines indicate the best--fit power--law and the $\pm$2$\sigma$ c.l. 
region of the correlation as determined by \cite{Amati08b}. The red dashed 
line is the best--fit power--law of the \fermi{} GRBs only.} 
\label{fig02} 
\end{figure}



%
\end{document}